\documentclass{article}
\usepackage{spconf}
\usepackage{amsmath}
\usepackage{graphicx}
\usepackage{cancel}
\usepackage{amssymb}
\usepackage{stmaryrd}
\usepackage{url}
\usepackage{hyperref}

\def\pyannote{{\small\texttt{pyannote}}}

\title{\texttt{pyannote.audio}: neural building blocks for speaker diarization}

\name{
\begin{tabular}{c}Herv\'{e} Bredin \; Ruiqing Yin \; Juan Manuel Coria \; Gregory Gelly \; Pavel Korshunov\thanks{This research was partly funded by the French National Research Agency (ANR) through the ODESSA (ANR-15-CE39-0010) and PLUMCOT (ANR-16-CE92-0025) projects. We would like to thank Claude Barras for providing the overlapped speech detection output corresponding to system $L_1$ in Table 2 of~\cite{Charlet2013}, Neville Ryant for the speaker diarization output of the winning submission to DIHARD 2019~\cite{Diez2019, dihard}, Marie Kune{\v{s}}ov{\'a} for the overlapped speech detection output corresponding to system \emph{"AMI test (all subsets) + dereverberation"} in Table 2 of~\cite{Kunesova2019}, and Sylvain Meignier for the speaker diarization output of~\cite{S4D} on ETAPE dataset.} \\
Marvin Lavechin \; Diego Fustes \; Hadrien Titeux \; Wassim Bouaziz \; Marie-Philippe Gill\end{tabular}}

\address{\texttt{github.com/pyannote/pyannote-audio}\vspace{-0.17cm}}

\begin{document}

\maketitle

\begin{abstract}
We introduce {\small \texttt{pyannote.audio}}, an open-source toolkit written in Python for speaker diarization.
Based on {\small \texttt{PyTorch}} machine learning framework, it provides a set of trainable end-to-end neural building blocks that can be combined and jointly optimized to build speaker diarization pipelines.
{\small \texttt{pyannote.audio}} also comes with pre-trained models covering a wide range of domains for voice activity detection, speaker change detection, overlapped speech detection, and speaker embedding -- reaching state-of-the-art performance for most of them.
\end{abstract}


\vspace{-0.17cm}
\section{Introduction}
\label{sec:intro}

\begin{figure*}[htb]
    \centering
    \includegraphics[width=0.8\linewidth]{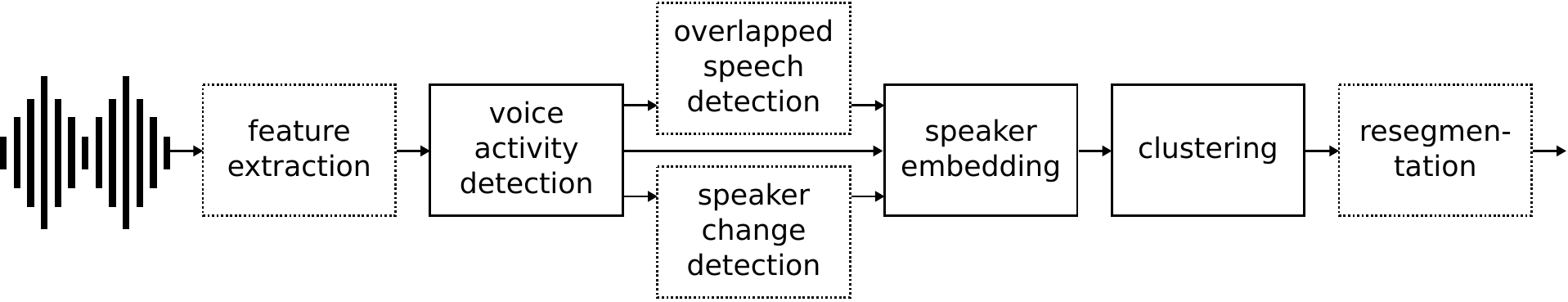}
    \caption{{\small \texttt{pyannote.audio}} provides a collection of modules that can be jointly optimized to build a speaker diarization pipeline.}
    \label{fig:pipeline}
\end{figure*}

Speaker diarization is the task of partitioning an audio stream into homogeneous temporal segments according to the identity of the speaker. As depicted in Figure~\ref{fig:pipeline}, this is usually addressed by putting together a collection of building blocks, each tackling a specific task (e.g. voice activity detection, clustering, or re-segmentation).

In this paper, we introduce {\small \texttt{pyannote.audio}}, an open-source toolkit written in {\small \texttt{Python}} and based on {\small \texttt{PyTorch}} machine learning framework, that provides end-to-end neural implementations for each of them.
A few open-source toolkits already exist that also address speaker diarization. Each one of them have its own pros and cons and we encourage the reader to build their own opinion by trying them:
\begin{description}
\vspace{-0.17cm}
\item[{\small \texttt{S4D}}] ({\small \texttt{SIDEKIT}} for diarization~\cite{S4D}) \emph{"provides an educational and efficient toolkit for speaker diarization including the whole chain of treatment"}. Among all {\small \texttt{pyannote.audio}} alternatives, it is the most similar: written in Python, it provides most of the aforementioned blocks, and goes all the way down to the actual evaluation of the system. However, it differs from {\small \texttt{pyannote.audio}} in its focus on traditional approaches (i.e. before the major shift towards deep learning) and the lack of joint optimization of the pipeline;
\vspace{-0.17cm}
\item[{\small \texttt{Kaldi}}] provides a few speaker diarization recipes but is not written in Python and is mostly dedicated to building speech and speaker recognition systems~\cite{kaldi};
\vspace{-0.17cm}
\item[{\small \texttt{ALIZ\'{E}}}] and its {\small \texttt{LIA\_SpkSeg}} extension for speaker diarization are written in C++ and do not provide recent deep learning approaches for speaker diarization~\cite{alize};
\vspace{-0.17cm}
\item[{\small \texttt{pyAudioAnalysis}}] is written in Python and addresses more general audio signal analysis, though it can be used for speaker diarization~\cite{pyAudioAnalysis}.
\end{description}

\vspace{-0.17cm}
\section{Feature extraction \\with built-in data augmentation}
\label{sec:feature_extraction}

While {\small\texttt{pyannote.audio}} supports training models from the waveform directly (e.g. using SincNet learnable features~\cite{Ravanelli2018}), the {\small\texttt{pyannote.audio.features}} module provides a collection of standard feature extraction techniques such as MFCCs or spectrograms using the implementation available in  the {\small\texttt{librosa}} library~\cite{librosa}. They all inherit from the same {\small\texttt{FeatureExtraction}} base class that supports on-the-fly data augmentation which is very convenient for training neural networks. For instance, it supports extracting features from random audio chunks while applying  additive noise from databases such as MUSAN~\cite{musan}. Contrary to other tools that generate in advance a fixed number of augmented versions of each original audio file, {\small\texttt{pyannote.audio}} generates a virtually infinite number of versions as the augmentation is done on-the-fly every time an audio chunk is processed.

\vspace{-0.17cm}
\section{Sequence labeling}
\label{sec:labeling}

{\small\texttt{pyannote.audio.labeling}} provides a unified framework to train (usually recurrent) neural networks for several speaker diarization sub-modules, including voice activity detection~\cite{Gelly2018}, speaker change detection~\cite{Yin2017}, overlapped speech detection~\cite{Bullock2020}, and even re-segmentation~\cite{Yin2018}.

\vspace{-0.17cm}
\subsection{Principle}
\label{ssec:principle}

\begin{figure*}[]
    \centering
    \includegraphics[width=0.8\linewidth]{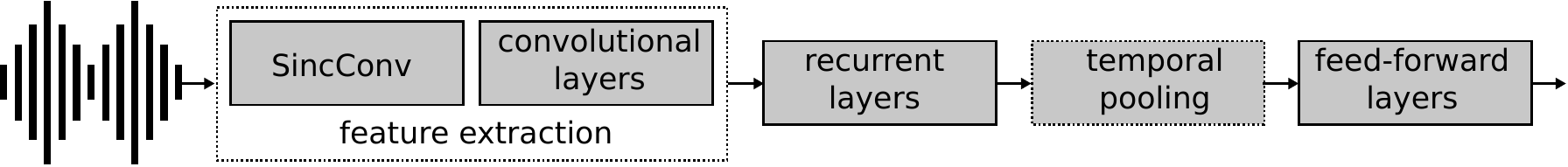}
    \caption{Generic {\small \texttt{PyanNet}} end-to-end architecture used for sequence labeling (without pooling) and embedding (with pooling).}
    \label{fig:pyannet}
\end{figure*}

Each of them can be addressed as a sequence labeling task where the input is the sequence of feature vectors $\mathbf{X} = \{\mathbf{x}_1, \mathbf{x}_2, \ldots, \mathbf{x}_T\}$ and the expected output is the corresponding sequence of labels $\mathbf{y} = \{y_1, y_2,\ldots, y_T\}$ with $y_t \in \llbracket 1; K \rrbracket$ where the number of classes $K$ depends on the task. {\small\texttt{pyannote.audio}} provides generic code to train a neural network $f: \mathbf{X}\rightarrow\mathbf{y}$ that matches a feature sequence $\mathbf{X}$ to the corresponding label sequence $\mathbf{y}$. The choice of the actual neural network architecture is left to the user, though {\small\texttt{pyannote.audio}} does provide pre-trained PyTorch models sharing the same generic PyanNet base architecture summarized in Figure~\ref{fig:pyannet}.

Because processing long audio files of variable lengths is neither practical nor efficient, {\small \texttt{pyannote.audio}} relies on shorter fixed-length sub-sequences. At training time, fixed-length sub-sequences are drawn randomly from the training set to form  mini-batches, increasing training samples variability (data augmentation) and training time (shorter sequences). At test time, audio files are processed using overlapping sliding windows of the same length as used in training. For each time step $t$, this results in several overlapping sequences of $K$-dimensional prediction scores, which are averaged to obtain the final score of each class.

\vspace{-0.17cm}
\subsection{Voice activity detection}
\label{ssec:vad}

Voice activity detection is the task of detecting speech regions in a given audio stream or recording. It can be addressed in {\small\texttt{pyannote.audio}} using the above principle with $K=2$: $y_{t} = 0$ if there is no speech at time step $t$ and $y_t = 1$ if there is. At test time, time steps with prediction scores greater than a tunable threshold $\theta_{\text{VAD}}$ are marked as speech. Overall, this essentially implements a simplified version of the voice activity detector originally described in~\cite{Gelly2018}. Pre-trained models are available, reaching state-of-the-art performance on a range of datasets, as reported in Table~\ref{tab:vad}.

\begin{table*}[htb]
    \centering
    \begin{tabular}{|l|r|rr|r|rr|r|rr|}
        \hline
            & \multicolumn{3}{c|}{\textbf{AMI}} & \multicolumn{3}{c|}{\textbf{DIHARD}} & \multicolumn{3}{c|}{\textbf{ETAPE}}\\
            & DetER & FA & Miss & DetER & FA & Miss & DetER & FA & Miss \\
        \hline
        Baseline & & & & 11.2 \cite{Diez2019, dihard} &  6.5 & 4.7 &  7.7 \cite{S4D} & 7.5 & 0.2 \\
        \hline
        \pyannote~(MFCC)~\cite{Gelly2018} & 6.3 \scriptsize{5.5} & 3.5 \scriptsize{3.1} & 2.7 \scriptsize{2.4} & 10.5 \scriptsize{10.0} & 6.8 \scriptsize{5.4} & 3.7 \scriptsize{4.6} & 5.6 \scriptsize{4.2} & 5.2 \scriptsize{3.6} & 0.4 \scriptsize{0.6} \\
        \pyannote~(waveform) & 6.0 \scriptsize{5.8} & 3.6 \scriptsize{3.4} & 2.4 \scriptsize{2.4} & 9.9 \scriptsize{9.3} & 5.7 \scriptsize{3.7} & 4.2 \scriptsize{5.6} & 4.9 \scriptsize{3.7} & 4.2 \scriptsize{2.9} & 0.7 \scriptsize{0.8}\\
        \hline

    \end{tabular}
    \caption{Evaluation of pre-trained voice activity detection models, in terms of detection error (DetER \%), false alarm (FA \%), and missed detection (Miss \%) rates. Results on the development set are reported using small font size.  We report two {\small{\texttt{pyannote.audio}}} variants: the first one is based on handcrafted features (MFCCs) and the other one is an end-to-end model processing the waveform directly.
    \emph{Baseline} corresponds to the best result we could find in the literature as of October 2019.}
    \label{tab:vad}
\end{table*}

\vspace{-0.17cm}
\subsection{Speaker change detection}

Speaker change detection is the task of detecting speaker change points in a given audio stream or recording. It can be addressed in {\small\texttt{pyannote.audio}} using the same sequence labeling principle with $K=2$: $y_{t} = 0$ if there is no speaker change at time step $t$ and $y_t = 1$ if there is. To address the class imbalance problem and account for human annotation imprecision, time steps $\left\{t \;| \; |t - t^*| < \delta\right\}$ in the close temporal neighborhood of a speaker change point $t^*$ are artificially labeled as positive for training. In practice, the order of magnitude of $\delta$ is $200$ms. At test time, time steps corresponding to prediction scores local maxima and greater than a tunable threshold $\theta_{\text{SCD}}$ are marked as speaker change points. Overall, this essentially implements a version of the speaker change detector originally described in~\cite{Yin2017}. Pre-trained models are available, reaching state-of-the-art performance on a range of datasets, as reported in Table~\ref{tab:scd}.

\begin{table*}[]
    \centering
    \begin{tabular}{|l|r|r|r|r|r|r|}
        \hline
        & \multicolumn{2}{c|}{\textbf{AMI}} & \multicolumn{2}{c|}{\textbf{DIHARD}} & \multicolumn{2}{c|}{\textbf{ETAPE}} \\
        & Purity & Coverage & Purity & Coverage & Purity & Coverage \\
        \hline
        Baseline &  &  &  &  & {\scriptsize{91.0}} \cite{Yin2017} & {\scriptsize{90.9}} \cite{Yin2017}\\
        \hline
        \pyannote~(MFCC)    & 89.4 \scriptsize{90.0} & 78.7 \scriptsize{75.2} & 92.4 \scriptsize{90.0} &  74.5 \scriptsize{76.6} & 90.1 \scriptsize{90.0} & 95.9 \scriptsize{95.7}   \\
        \pyannote~(waveform) & 90.4 \scriptsize{90.0} & 84.2 \scriptsize{83.5} & 86.8 \scriptsize{84.5} & 93.7 \scriptsize{93.4} & 89.3 \scriptsize{90.0} & 97.2 \scriptsize{98.2}  \\
        \hline

    \end{tabular}
    \caption{Evaluation of pre-trained speaker change detection models, in terms of speech turn coverage (\%) and purity (\%). Results on the development set are reported using small font size. We report two {\small{\texttt{pyannote.audio}}} variants: the first one is based on handcrafted features (MFCCs) and the other one is an end-to-end model processing the waveform directly.
    \emph{Baseline} corresponds to the best result we could find in the literature as of October 2019.}
    \label{tab:scd}
\end{table*}

\vspace{-0.17cm}
\subsection{Overlapped speech detection}
\label{ssec:ovl}

\begin{figure}
    \centering
    \includegraphics[width=0.75\linewidth]{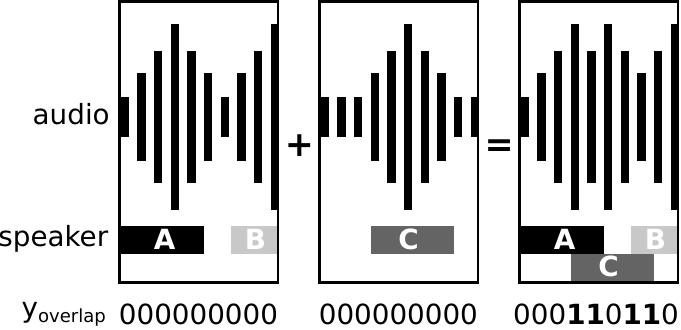}
    \caption{To increase the number of positive training samples for overlapped speech detection, {\small \texttt{pyannote.audio}} creates artificial samples by summing two random audio chunks.}
    \label{fig:ovl}
\end{figure}

Overlapped speech detection is the task of detecting regions where at least two speakers are speaking at the same time. It is addressed in {\small\texttt{pyannote.audio}} using the same sequence labeling principle with $K=2$: $y_{t} = 0$ if there is zero or one speaker at time step $t$ and $y_t = 1$ if there are two speakers or more. To address the class imbalance problem, half of the training sub-sequences are artificially made of the weighted sum of two random sub-sequences, as depicted in Figure~\ref{fig:ovl}. At test time, time steps with prediction scores greater than a tunable threshold $\theta_{\text{OSD}}$ are marked as overlapped speech. Pre-trained models are available, reaching state-of-the-art performance on a range of datasets, as reported in Table~\ref{tab:ovl}.

\begin{table*}[htb]
    \centering
    \begin{tabular}{|l|r|r|r|r|r|r|}
        \hline
        & \multicolumn{2}{c|}{\textbf{AMI}} & \multicolumn{2}{c|}{\textbf{DIHARD}} & \multicolumn{2}{c|}{\textbf{ETAPE}}\\
        & Precision & Recall & Precision & Recall & Precision & Recall \\
        \hline
        Baseline                    & 75.8 {\scriptsize{80.5}} \cite{Kunesova2019} & 44.6 {\scriptsize{50.2}} \cite{Kunesova2019} & & & 60.3 \cite{Charlet2013} &  52.7 \cite{Charlet2013} \\
                \hline
        \pyannote~(MFCC) &  91.9 \scriptsize{90.0} & 48.4 \scriptsize{52.5} & 58.0 \scriptsize{73.8} & 17.6 \scriptsize{14.0} & 67.1 \scriptsize{55.0} & 57.3 \scriptsize{55.3} \\
        \pyannote~(waveform) & 86.8 \scriptsize{90.0} & 65.8 \scriptsize{63.8} & 64.5 \scriptsize{75.3} & 26.7 \scriptsize{24.4} & 69.6 \scriptsize{60.0} & 61.7 \scriptsize{63.6} \\
\hline

    \end{tabular}
    \caption{Evaluation of pre-trained overlapped speech detection models, in terms of precision (\%) and recall (\%). Results on the development set are reported using small font size. We report two {\small{\texttt{pyannote.audio}}} variants: the first one is based on handcrafted features (MFCCs) and the other one is an end-to-end model processing the waveform directly.
    \emph{Baseline} corresponds to the best result we could find in the literature as of October 2019.}
    \label{tab:ovl}
\end{table*}

\vspace{-0.17cm}
\subsection{Re-segmentation}
\label{ssec:reseg}

Re-segmentation is the task of refining speech turns boundaries and labels coming out of a diarization pipeline.
Though it is an unsupervised task, it is addressed in {\small\texttt{pyannote.audio}} using the same sequence labeling principle with $K=\kappa + 1$ where $\kappa$ is the number of speakers hypothesized by the diarization pipeline: $y_t = 0$ if no speaker is active at time step $t$ and $y_t = k$ if speaker $k \in \llbracket 1; \kappa \rrbracket$ is active. Because the re-segmentation step is unsupervised by design, one cannot pre-train re-segmentation models.
For each audio file, a new re-segmentation model is trained from scratch using the (automatic, hence imperfect) output of a diarization pipeline as training labels. Once trained for a number of epochs $\epsilon$ (one epoch being one complete pass on the file), the model is applied on the very same file it was trained from -- making this approach completely unsupervised. Each time step is assigned to the class (non-speech or one of the $\kappa$ speakers) with highest prediction scores. This essentially implements a version of the re-segmentation approach originally described in~\cite{Yin2018} where it was found that $\epsilon = 20$ is a reasonable number of epochs. This re-segmentation step may be extended to also assign the class with the second highest prediction score to overlapped speech regions~\cite{Bullock2020}. As reported in Table~\ref{tab:dia}, this may lead to significant performance improvement.

\vspace{-0.17cm}
\section{Sequence embedding and clustering}
\label{sec:clustering}

Like voice activity detection, clustering is one of the most important part of any speaker diarization pipeline.
It consists in grouping speech segments according to the actual identity of the speaker. As of October 2019, most best performing speaker diarization systems rely on \emph{x-vectors} as input to the clustering step. They are usually extracted from a fixed-length sliding window, and the pairwise similarity matrix is obtained with probabilistic linear discriminant analysis (PLDA).
There exist plenty of open-source implementations of \emph{x-vectors} and PLDA already~\cite{S4D, kaldi}, and {\small\texttt{pyannote.audio}} does not provide yet another version of those approaches.

Instead, the clustering step is simplified by using metric learning approaches to train speaker embedding that are directly optimized for a predefined (usually cosine) distance, reducing the need for techniques like PLDA\footnote{\href{https://www.clsp.jhu.edu/faculty/jesus-villalba/}{x-vector} \href{https://www.clsp.jhu.edu/faculty/paola-garcia/}{\emph{aficionados}} would still suggest to use PLDA anyway...}. As reported in Table~\ref{tab:emb}, a growing collection of metric learning approaches are implemented in {\small \texttt{pyannote.audio.embedding}} that provides a unified framework for this family of approaches.

\begin{table}[]
    \centering
    \begin{tabular}{|l|r|}
        \hline
        Baseline & 3.1 \cite{x-vector} \\
        \hline
        \pyannote~triplet loss & 7.0 {\scriptsize{WIP}} \\
        \pyannote~additive angular margin loss & 10.0 {\scriptsize{WIP}} \\
        \pyannote~constrastive loss & 17.4 {\scriptsize{WIP}} \\
        \pyannote~center loss & 16.6 {\scriptsize{WIP}} \\
        \pyannote~congenerous cosine loss& 18.4 {\scriptsize{WIP}} \\
        \hline
    \end{tabular}
    \caption{Evaluation of speaker embedding on VoxCeleb 1 speaker verification task, in terms of equal error rate (\%). {\scriptsize{WIP}} indicates that training is still ongoing.}
    \label{tab:emb}
\end{table}

Like other blocks, speaker embeddings can be trained either from handcrafted features, or from the waveform directly in an end-to-end manner. Though pre-trained end-to-end models do not reach state-of-the-art performance on VoxCeleb speaker verification task, Table~\ref{tab:dia} shows that they do lead to state-of-the-art performance for some speaker diarization test sets.

\vspace{-0.17cm}
\section{Tunable pipelines}
\label{sec:pipeline}

While each building block has to be trained separately, {\small \texttt{pyannote.audio.pipeline}} combines them into a speaker diarization pipeline whose hyper-parameters are optimized jointly to minimize the diarization error rate (or any other metric available in {\small \texttt{pyannote.metrics}}~\cite{pyannote.metrics}). As discussed in~\cite{Yin2018}, this joint optimization process usually leads to better results than the late combination of multiple building blocks that were tuned independently from each other. Table~\ref{tab:dia} reports the results of the pipeline we introduced in~\cite{Yin2018} for which a bunch of hyper-parameters where jointly optimized, including $\theta_\text{VAD}$ for voice activity detection and $\theta_\text{SCD}$ for speaker change detection. We also report the improvement brought by the integration of the overlapped speech detection in the re-segmentation step (introduced in Section~\ref{ssec:reseg} and further described in~\cite{Bullock2020}).

\begin{table}[htb]
    \centering
    \begin{tabular}{|l|r|r|r|}
        \hline
        & \multicolumn{1}{c|}{\textbf{AMI}} & \multicolumn{1}{c|}{\textbf{DIHARD}} & \multicolumn{1}{c|}{\textbf{ETAPE}} \\
        \hline
        Baseline & \cancel{8.4} \cite{Maciejewski2018} &  27.1 \cite{Diez2019} &  24.5 \cite{S4D}\\
        \hline
        \pyannote~\cite{Yin2018} & \cancel{4.6} 29.6 \scriptsize{29.7} & 34.4 \scriptsize{31.5} & 24.0 \scriptsize{18.3} \\
        ... with overlap~\cite{Bullock2020} & 24.8 \scriptsize{24.7} &  &  \\
        \hline
    \end{tabular}
    \caption{Evaluation of speaker diarization pipelines in terms of diarization error rate (\%). \emph{Baseline} corresponds to the best result we could find in the literature as of October 2019. Since \cite{Maciejewski2018}~only reports confusion error rates (assuming oracle speech activity detection, omitting overlapped speech regions, and using a collar), those are marked as crossed-out \cancel{numbers}.}
    \label{tab:dia}
\end{table}

\vspace{-0.17cm}
\section{Reproducible results}
\label{sec:reproducible}

{\small \texttt{pyannote.audio}} provides a set of command line tools for training, validation, and application of modules listed in Figure~\ref{fig:pipeline}. Reproducible research is facilitated by the systematic use of {\small {\texttt{pyannote.metrics}}}~\cite{pyannote.metrics} and configuration files, while strict enforcement of train/dev/eval split with {\small {\texttt{pyannote.database}}} ensures machine learning good practices.

It also comes with a collection of pre-trained models whose performance has already been reported throughout the paper in Table~\ref{tab:vad} for voice activity detection, Table~\ref{tab:scd} for speaker change detection, Table~\ref{tab:ovl} for overlapped speech detection, and Table~\ref{tab:emb} for speaker embedding. While speaker embeddings were trained and tested on VoxCeleb~\cite{voxceleb}, all other models (including the full diarization pipeline) were trained, tuned, and tested on three different datasets, covering a wide range of domains: meetings for AMI~\cite{ami}, broadcast news for ETAPE~\cite{etape}, and up to 11 different domains for DIHARD~\cite{dihard}.

They all rely on the same generic \emph{PyanNet} architecture which is depicted in Figure~\ref{fig:pyannet}. Sequence labeling tasks were trained on 2s audio chunks, either with handcrafted MFCC features (19 coefficients extracted every 10ms on 25ms windows, with first- and second-order derivatives) or with trainable SincNet features (using the configuration of the original paper~\cite{Ravanelli2018}). The end-to-end variant consistently outperforms the one based on handcrafted features for all tasks but speaker embedding (for which we have yet to train the handcrafted features variant), defining a new state-of-the-art performance for most cases.

The rest of the network includes two stacked bi-directional LSTM recurrent layers (each with 128 units in both forward and backward directions), no temporal pooling, two feed-forward layers (128 units, \emph{tanh} activation) and a final classification layer (2 units, \emph{softmax} activation). Speaker embedding uses a slightly wider (512 units instead of 128) and deeper (3 recurrent layers instead of 2) network, relies on \emph{x-vector}-like statistical temporal pooling~\cite{x-vector} and has been trained on shorter 500ms audio chunks. More details on the training process (number of epochs, optimizer, etc.) can be found directly in the associated configuration files, available from {\small \texttt{pyannote.audio}} repository.

\newpage

\bibliographystyle{IEEEbib}
\bibliography{refs}

\begin{thebibliography}{10}

\bibitem{Charlet2013}
Delphine Charlet, Claude Barras, and Jean-Sylvain Li\'{e}nard,
\newblock ``Impact of overlapping speech detection on speaker diarization for
  broadcast news and debates,''
\newblock in {\em Proc. ICASSP 2013}, May 2013.

\bibitem{Diez2019}
Mireia Diez, Luk\'{a}š Burget, Shuai Wang, Johan Rohdin, and Jan Černocký,
\newblock ``{Bayesian HMM Based x-Vector Clustering for Speaker Diarization},''
\newblock in {\em Proc. Interspeech 2019}, 2019.

\bibitem{dihard}
Neville Ryant, Kenneth Church, Christopher Cieri, Alejandrina Cristia, Jun Du,
  Sriram Ganapathy, and Mark Liberman,
\newblock ``{The Second DIHARD Diarization Challenge: Dataset, Task, and
  Baselines},''
\newblock in {\em Proc. Interspeech 2019}, 2019, pp. 978--982.

\bibitem{Kunesova2019}
Marie Kune{\v{s}}ov{\'a}, Marek Hr{\'u}z, Zbyn{\v{e}}k Zaj{\'i}c, and Vlasta
  Radov{\'a},
\newblock ``{Detection of Overlapping Speech for the Purposes of Speaker
  Diarization},''
\newblock in {\em Speech and Computer}, 2019, pp. 247--257.

\bibitem{S4D}
Pierre-Alexandre Broux, Florent Desnous, Anthony Larcher, Simon Petitrenaud,
  Jean Carrive, and Sylvain Meignier,
\newblock ``{S4D: Speaker Diarization Toolkit in Python},''
\newblock in {\em Proc. Interspeech 2018}, 2018, pp. 1368--1372.

\bibitem{kaldi}
Daniel Povey, Arnab Ghoshal, Gilles Boulianne, Lukas Burget, Ondrej Glembek,
  Nagendra Goel, Mirko Hannemann, Petr Motlicek, Yanmin Qian, Petr Schwarz, Jan
  Silovsky, Georg Stemmer, and Karel Vesely,
\newblock ``The kaldi speech recognition toolkit,''
\newblock in {\em IEEE 2011 Workshop on Automatic Speech Recognition and
  Understanding}. Dec. 2011, IEEE Signal Processing Society,
\newblock IEEE Catalog No.: CFP11SRW-USB.

\bibitem{alize}
Jean-Fran\c{c}ois Bonaster, F~Wils, and Sylvain Meignier,
\newblock ``{Aliz\'e, a free Toolkit for Speaker Recognition},''
\newblock in {\em {Proc. Interspeech 2005}}, 2005.

\bibitem{pyAudioAnalysis}
Theodoros Giannakopoulos,
\newblock ``pyaudioanalysis: An open-source python library for audio signal
  analysis,''
\newblock {\em PloS one}, vol. 10, no. 12, 2015.

\bibitem{Ravanelli2018}
Mirco Ravanelli and Yoshua Bengio,
\newblock ``Speaker recognition from raw waveform with sincnet,''
\newblock in {\em Proc. SLT 2018}, 2018.

\bibitem{librosa}
Brian McFee, Vincent Lostanlen, Matt McVicar, Alexandros Metsai, Stefan Balke,
  Carl Thomé, Colin Raffel, Dana Lee, Kyungyun Lee, Oriol Nieto, Jack Mason,
  Frank Zalkow, Dan Ellis, Eric Battenberg, Виктор Морозов,
  Ryuichi Yamamoto, Rachel Bittner, Keunwoo Choi, Josh Moore, Ziyao Wei,
  nullmightybofo, Pius Friesch, Fabian-Robert Stöter, Darío Hereñú,
  Thassilo, Taewoon Kim, Matt Vollrath, Adam Weiss, CJ~Carr, and ajweiss dd,
\newblock ``librosa/librosa: 0.7.0,'' July 2019.

\bibitem{musan}
David Snyder, Guoguo Chen, and Daniel Povey,
\newblock ``{MUSAN}: {A} {M}usic, {S}peech, and {N}oise {C}orpus,'' 2015.

\bibitem{Gelly2018}
Gregory Gelly and Jean-Luc Gauvain,
\newblock ``{Optimization of RNN-Based Speech Activity Detection},''
\newblock {\em IEEE/ACM Transactions on Audio, Speech, and Language
  Processing}, vol. 26, no. 3, pp. 646--656, March 2018.

\bibitem{Yin2017}
Ruiqing Yin, Herv\'{e} Bredin, and Claude Barras,
\newblock ``{Speaker Change Detection in Broadcast TV Using Bidirectional Long
  Short-Term Memory Networks},''
\newblock in {\em Proc. Interspeech 2017}, 2017.

\bibitem{Bullock2020}
Latan\'{e} Bullock, Herv\'e Bredin, and Leibny~Paola Garcia-Perera,
\newblock ``{Overlap-Aware Resegmentation for Speaker Diarization},''
\newblock Submitted to ICASSP 2020.

\bibitem{Yin2018}
Ruiqing Yin, Herv\'{e} Bredin, and Claude Barras,
\newblock ``{Neural Speech Turn Segmentation and Affinity Propagation for
  Speaker Diarization},''
\newblock in {\em Proc. Interspeech 2018}, 2018, pp. 1393--1397.

\bibitem{x-vector}
David Snyder, Daniel Garcia-Romero, Gregory Sell, Daniel Povey, and Sanjeev
  Khudanpur,
\newblock ``{X-Vectors: Robust DNN Embeddings for Speaker Recognition},''
\newblock in {\em Proc. ICASSP 2018}, 2018.

\bibitem{pyannote.metrics}
Herv\'e Bredin,
\newblock ``{pyannote.metrics: a toolkit for reproducible evaluation,
  diagnostic, and error analysis of speaker diarization systems},''
\newblock in {\em {Proc. Interspeech 2017}}, Stockholm, Sweden, August 2017.

\bibitem{Maciejewski2018}
M.~{Maciejewski}, David {Snyder}, V.~{Manohar}, Najim {Dehak}, and Sanjeev
  {Khudanpur},
\newblock ``{Characterizing Performance of Speaker Diarization Systems on
  Far-Field Speech Using Standard Methods},''
\newblock in {\em Proc. ICASSP 2018}, 2018.

\bibitem{voxceleb}
Joon~Son Chung, Arsha Nagrani, and Andrew Zisserman,
\newblock ``Voxceleb2: Deep speaker recognition,''
\newblock in {\em Proc. InterSpeech 2018}, 2018.

\bibitem{ami}
Jean Carletta,
\newblock ``{Unleashing the killer corpus: experiences in creating the
  multi-everything AMI Meeting Corpus},''
\newblock {\em Language Resources and Evaluation}, vol. 41, no. 2, 2007.

\bibitem{etape}
Guillaume Gravier, Gilles Adda, Niklas Paulson, Matthieu Carr{\'e}, Aude
  Giraudel, and Olivier Galibert,
\newblock ``{The ETAPE Corpus for the Evaluation of Speech-based TV Content
  Processing in the French Language},''
\newblock in {\em {Proc. LREC 2012}}, 2012.

\end{thebibliography}

\end{document}